\documentclass[a4paper,11pt]{article}
\usepackage{pos}

\title{Stereo performance of SST-1M at different altitudes}

\author*[a]{Patrik \v Cechvala}
\author[a,b]{Vladim\'ir Novotn\'y}
\author[a]{Ji\v r\'i Bla\v zek}
\author[a]{Ana Laura M\"uller}
\author[a]{Jakub Jury\v sek}
\author[a]{Thomas Tavernier}
\onbehalf{on behalf of the SST-1M collaboration}

\affiliation[a]{FZU - Institute of Physics of the Czech Academy of Sciences, \\
Na Slovance 1999/2, Prague 8, Czech Republic}

\affiliation[b]{Institute of Particle and Nuclear Physics, Faculty of Mathematics and Physics, Charles~University,\\
  V Hole\v sovi\v ck\'ach 2, 180 00 Prague, Czech Republic}

\emailAdd{cechvala@fzu.cz}
\emailAdd{vladimir.novotny@matfyz.cuni.cz}
\emailAdd{blazekj@fzu.cz}
\emailAdd{mulleral@fzu.cz}
\emailAdd{jurysek@fzu.cz}
\emailAdd{tavernier@fzu.cz}

\abstract{The SST-1M telescopes are a pair of Imaging Atmospheric Cherenkov Telescopes (IACTs) that have been operating at the Ond\v rejov Observatory (510\,m a.s.l.) in the Czech Republic since 2022. Optimized for detecting gamma rays in the energy range 1-300\,TeV, they are capable of performing both mono and stereo observations. Despite challenging atmospheric conditions, SST-1M has successfully detected several galactic and extragalactic gamma-ray sources with energies reaching up to 200\,TeV during its ongoing commissioning.
In this study, we analyze the performance of the SST-1M telescopes at different locations to assess the impact of altitude and relative telescope spacing on their physics performance. The low-altitude site at 510\,m a.s.l. has already been investigated using both Monte Carlo simulations and real data. For comparison, we selected an intermediate-altitude site at 1420\,m a.s.l. corresponding to Pampa Amarilla in Argentina and a high-altitude site at 4270\,m a.s.l. corresponding to Hanle in India – both of which offer favorable astronomical conditions.}

\ConferenceLogo{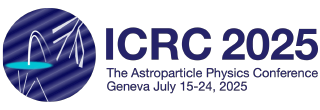}

\FullConference{39th International Cosmic Ray Conference (ICRC2025)\\
 15–24 July 2025\\
Geneva, Switzerland\\}

\begin{document}
\maketitle

\section{Introduction}
Imaging Atmospheric Cherenkov Telescopes (IACTs) represent an important ground-based detection method, enabling the observation of gamma-ray astrophysical sources with energies exceeding several tens of GeV. The Single-Mirror Small-Size Telescope (SST-1M) is an IACT primarily focused on detecting gamma rays within an energy range from approximately 1 TeV up to several hundreds of TeV. The SST-1M was developed by a consortium of institutes from the Czech Republic, Poland, and Switzerland as a prototype for Cherenkov Telescope Array Observatory (CTAO). Two prototypes of SST-1M were constructed and, in 2022 installed at the Ond\v rejov Observatory in the Czech Republic at an altitude of 510\,m above sea level (a.s.l.). A detailed description of the telescopes and their operation in Ond\v rejov can be found in~\cite{SST-1M}.
Separated by 155.2\,m, the telescopes can perform stereo observations of extensive air showers (EAS) induced by gamma rays. The SST-1M mini-array has successfully undergone commissioning and detected several galactic and extragalactic astrophysical sources. The first scientific results of the observation of the Crab Nebula were recently published \cite{Alispach_2025b}, confirming the telescope`s expected capabilities. For this publication, we will refer to each telescope with its specific label: SST-1M-1 (long $= 14.782924^{\circ}$, lat $= 49.911800^{\circ}$) and SST-1M-2 (long $= 14.782078^{\circ}$, lat $= 49.913084^{\circ}$). This notation follows that established in~\cite{Alispach_2025b}.\\
Although SST-1M telescopes have proven capabilities of detecting gamma-ray showers from a low-altitude site \cite{Alispach_2025b}, the atmospheric conditions in Ond\v rejov are not satisfactory. Due to the low altitude, the SST-1M telescopes exhibit the best performance at the highest energies. The Ond\v rejov site offers in total $\sim$ 3300 \,h/year\footnote{\url{https://stel.asu.cas.cz/en/telescope/observations/}} of dark hours with the Sun more than 12$^{\circ}$ below the horizon (corresponding to 100\% time between nautical twilights). This estimate of observation hours does not account for bad weather conditions or nights with increased moonlight. After the first years of observations in Ond\v rejov, the amount of achievable observation hours is $\sim$ 330\,h/year in stereo mode and $\sim$ 250\,h/year after quality selection. One can compare this with the current generation of IACTs (H.E.S.S. $\sim$ 1500\,h/year\footnote{\url{https://www.mpi-hd.mpg.de/HESS/pages/home/som/2022/07/}}, MAGIC $\sim$ 1400\,h/year\footnote{\url{https://magic.mpp.mpg.de/public/magicop/}}, VERITAS $\sim$ 1200\,h/year \cite{veritas_hours}). Given these considerations, there is currently an ongoing discussion about relocating the telescopes to a more favorable site where they can achieve their full potential. Two possible future sites have been chosen. To provide scientific arguments for a proper and responsible decision regarding the preferred site, we computed the performance of the SST-1M telescopes at each site using the standard SST-1M analysis procedure. We discuss the chosen telescope layout and describe the simulations. We focus on the scientific performance of the telescopes, extraction of the Instrument Response Functions (IRF), and their comparison between the considered sites.
\section{Methods}

\subsection{Studied sites and simulated telescope layout}
\label{sec:layout}

A detailed study~\cite{cta_site_selection} on the performance of IACTs at different sites has been conducted for the purposes of CTAO. This study showed that the optimal IACT performance can be achieved at intermediate altitudes between approximately 1.6\,km and 2.5\,km a.s.l., due to an optimal interplay between the attenuation of Cherenkov light and the development of the EAS.\\
For our analysis, two sites have been selected as propositions for the future placement of the SST-1M telescopes. The first is Malarg\"ue in the province of Mendoza, Argentina, hereafter referred as Malarg\"ue. This site provides an intermediate altitudes of 1420\,m a.s.l. near the Andes mountains. The selected geographical coordinates of the telescopes used in Monte Carlo (MC) simulations were long $= -69.202001^{\circ}$, lat $= -35.150943^{\circ}$ for SST-1M-1 and long $= -69.200858^{\circ}$, lat $= -35.150404^{\circ}$ for SST-1M-2. These coordinates result in a spacing 120\,m between the telescopes in the ground frame. Due to the simulation settings during which the telescopes were pointing to the zenith angle of 20$^{\circ}$, this spacing is modified in the tilted frame. We define this as the effective spacing, which corresponds to 118.24\,m for the Malarg\"ue site.
The second considered site is Hanle in the south-eastern Ladakh union territory of India, hereafter referred as Hanle. This site is at significantly higher altitude of 4270\,m a.s.l. in Himalayas. The selected geographical coordinates for the telescopes used in the MC simulations were long $= 78.978956^{\circ}$, lat $= 32.780496^{\circ}$ for SST-1M-1 and long $= 78.979550^{\circ}$, lat $= 32.780208^{\circ}$ for SST-1M-2. The layout was scaled to match the telescope spacing in Malargüe, following the geometry of the Cherenkov light cone and assuming the shower maximum at an altitude of 7.5\,km a.s.l. corresponding to an energy of primary gamma ray $\sim$ 5\,TeV. The simulated spacing in the ground frame was 64\,m which corresponds to an effective spacing 63.01\,m in the tilted frame.

\subsection{Monte Carlo simulations}
The sets of MC simulations for each considered site were prepared. MC simulations of particle showers, including the production of Cherenkov light, were made in \texttt{CORSIKA v7.7402}~\cite{1998cmcc.book.....H} using hadronic interaction models for low- and high-energy interactions \texttt{UrQMD}~\cite{urqmd2} and \texttt{QGSJet\,II-04}~\cite{qgs}, respectively.
After that, the attenuation of the light and the response of the SST-1M telescopes were simulated in \texttt{sim\_telarray v2021-12-25}~\cite{BERNLOHR2008149}.
The atmospheric density profile, together with the model of light attenuation in the atmosphere and the geomagnetic field, was set according to the investigated sites of Malarg\"ue and Hanle.
The SST-1M simulations were performed using the NSB rate of 55\,MHz and 72\,MHz for SST-1M-1 and SST-1M-2, respectively, and the camera response was adjusted correspondingly \cite{Alispach_2025b}.
The different rates were set due to slightly different hardware parameters of the two telescopes.

The telescopes pointed to 20$^{\circ}$ zenith angle, while the azimuth was set to $0^{\circ}$ and $180^{\circ}$ for the Malarg\"ue and Hanle sites, respectively, to cover the culmination of sources.
The effective distances of the two SST-1Ms were chosen to be 63.01\,m and 118.24\,m for Hanle and Malarg\"ue, respectively, as described in Section~\ref{sec:layout}. For reference, the effective distance for Ond\v rejov is 147.31\,m.
A set of training simulations of diffuse protons and gamma rays was used to optimize the reconstruction for each site.
Details about the Random Forest (RF) reconstruction procedure, together with the necessary amount of simulations, are described in~\cite{Alispach_2025b}. The analysis was performed using \texttt{sst1mpipe} \footnote{\url{https://github.com/SST-1M-collaboration/sst1mpipe}} ~\cite{sst1mpipe, jurysek_ICRC_proceeding, Alispach_2025b} software which was specifically developed for the low-level analysis of data and MC simulations of SST-1M telescopes. The version of the pipeline \texttt{v0.7.4} was used for the presented analysis. Althought, some minor modifications were implemented to account for the different observation sites, telescope layouts or telescope pointings.



\section{Performance of SST-1M telescopes at selected sites}

For each site, we followed the analysis steps and definitions for the analyzed parameters as described in~\cite{Alispach_2025b}. The perfomance for Malarg\"ue and Hanle are compared with those from Ond\v rejov, which serves as the reference site. 

\begin{figure}[h]
\centering
\includegraphics[width=.46\linewidth]{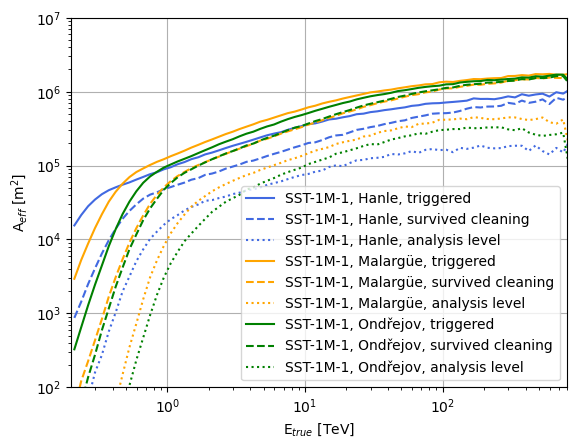}\hfill
\includegraphics[width=.46\linewidth]{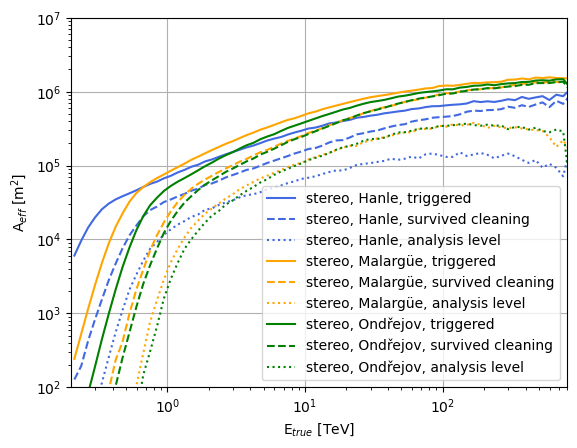}\hfill
\caption{Effective area for different sites: Hanle (4270\,m a.s.l., blue), Malarg\"ue (1420\,m a.s.l., orange), Ond\v rejov (510\,m a.s.l., green). \textit{Left}: mono (only for SST-1M-1), \textit{Right}: stereo. Different stages of the analysis are shown: triggered events (solid line), events that survived cleaning (dashed line) and events that survived quality, $\theta^{2}$ and gammaness cuts (dotted line). Different stages of the analysis are described in~\cite{Alispach_2025b}.}
\label{fig:aeff}
\end{figure}

The effective area for mono and stereo observations at different stages of the analysis is shown in Figure~\ref{fig:aeff}. The same quality cuts were applied as in~\cite{Alispach_2025b}.  Optimization of cuts on MC was performed to achieve the best detection significance for a source with an energy spectrum corresponding to Crab Nebula  ~\cite{crab_nebula_spectrum}. As expected, the detection capabilities for low-energy region gradually improve with increasing altitude. This enhancement, however, comes at the expense of loss of detection sensitivity at higher energies. For Malarg\"ue, larger effective area is observed across the whole energy region for mono mode. In stereo, there is a marginal loss at high energies, with the effective area almost matching that of Ond\v rejov and showing a minor gain at energies $\lesssim$ 15\,TeV at the analysis level. However, for the Hanle site, which is located at significantly higher altitude, this effect is clearly pronounced both in mono and stereo mode. At the analysis level, a gain in effective area is observed for energies $\lesssim$ 3\,TeV both for mono and stereo, accompanied by a significant suppression at higher energies for this site. For an energy of 100\,TeV the effective area for Hanle in stereo is $\sim$ 2.6 times smaller than the corresponding area for Ond\v rejov. For Malarg\"ue, the loss is on the order of $\sim$ 2.4\% compared to Ond\v rejov. At the energy domain of 10\,TeV the area for Hanle is $\sim$ 1.7 times smaller, while Malarg\"ue already shows a gain in effective area $\sim$ 2.3\%. At the energies around 1\,TeV, the improvement in detection capabilities with increasing altitude is clearly visible, with effective area $\sim$ 1.7 times larger for Malarg\"ue and $\sim$ 4.7 times larger for Hanle compared to Ond\v rejov. This difference increases as the energy decreases reaching for the Hanle site $\sim$ 81.9 times larger effective area at 600\,GeV. For the Malarg\"ue site, the effective area is approximately 2.5 times larger.          

\begin{figure}[h]
\includegraphics[width=.43\linewidth]{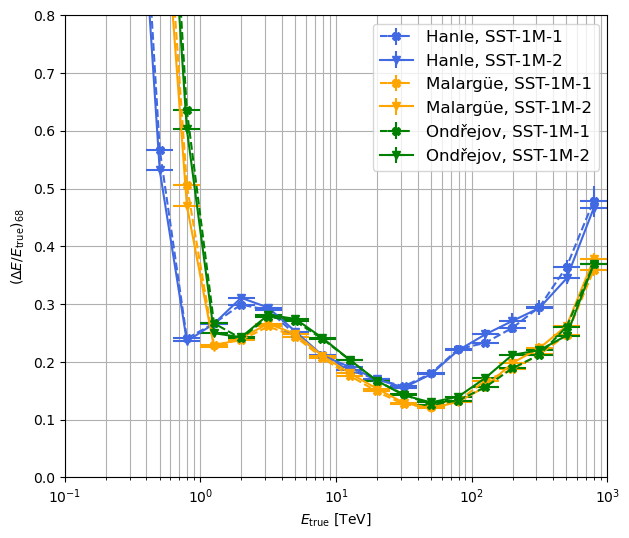}\hfill
\includegraphics[width=.43\linewidth]{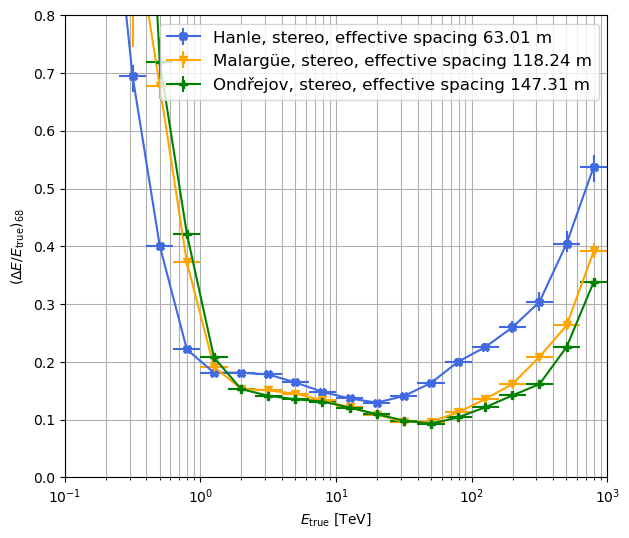}
\caption{Energy resolution for different sites: Hanle (4270\,m a.s.l., blue), Malarg\"ue (1420\,m a.s.l., orange), Ond\v rejov (510\,m a.s.l., green). \textit{Left}: mono (SST-1M-1 dashed line, SST-1M-2 solid line), \textit{Right}: stereo.}
\label{fig:ene}
\end{figure}

\begin{figure}[h]
\includegraphics[width=.45\linewidth]{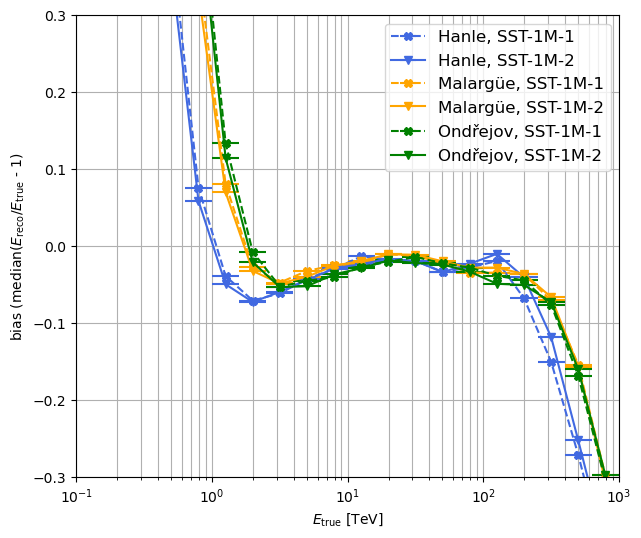}\hfill
\includegraphics[width=.45\linewidth]{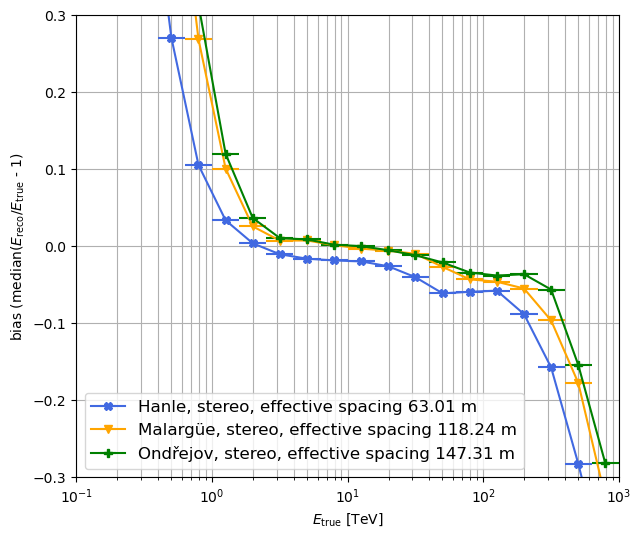}
\caption{Energy bias for different sites: Hanle (4270\,m a.s.l., blue), Malarg\"ue (1420\,m a.s.l., orange), Ond\v rejov (510\,m a.s.l., green). \textit{Left}: mono (SST-1M-1 dashed line, SST-1M-2 solid line), \textit{Right}: stereo.}
\label{fig:bias}
\end{figure}

Energy resolution and bias for both mono and stereo modes are shown in Figure~\ref{fig:ene} and Figure~\ref{fig:bias}, respectively. For mono observation, an improvement in energy resolution is observed at the Malarg\"ue site for energies $\lesssim$ 50\,TeV, with comparable values for higher energies. At Hanle, the energy resolution degrades almost over the entire studied energy range, except for the energies lower than $\sim$ 0.8\,TeV where it surpasses the resolution at Malarg\"ue. We observe that stereo observation improves energy resolution and bias for all 3 sites especially at energies below 10\,TeV. Stereo mode reflects the same trend as discussed above for the effective area, where energy resolution degrades at high energies due to the increasing altitude and improves at lowest energy bins. Degradation of the energy resolution at Malarg\"ue  begins to be visible for energies $\gtrsim$ 50\,TeV compared to Ond\v rejov. More significant suppression can be seen for Hanle site, already for energies $\gtrsim$ 30\,TeV. Conversely, an improvement is observed at the energies $\lesssim$ 1.3\,TeV for both analyzed sites, which is more distinctive for Hanle. The energy resolution in stereo mode reaches at Hanle $\sim$ 13\% at 20\,TeV. At Malarg\"ue, it is $\sim$ 9.6\% at 30\,TeV.

\begin{figure}[h]
\includegraphics[width=.43\linewidth]{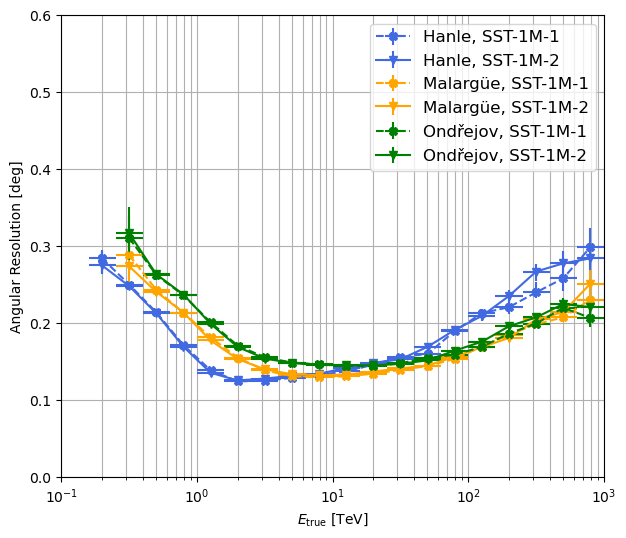}\hfill
\includegraphics[width=.43\linewidth]{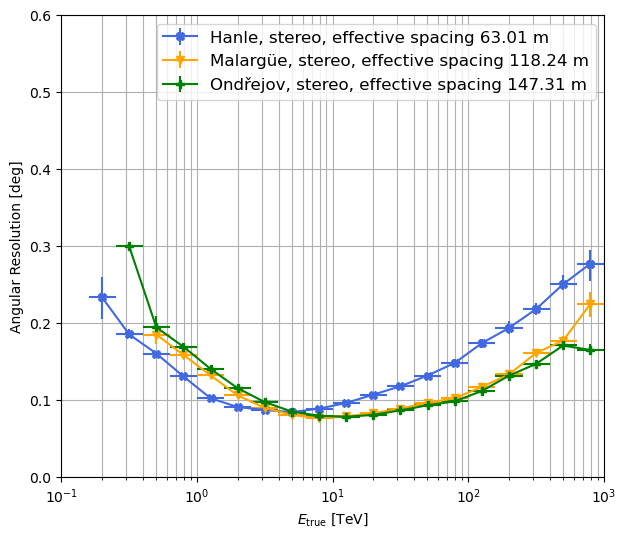}
\caption{Angular resolution for different sites: Hanle (4270 m a.s.l., blue), Malargüe (1420 m a.s.l., orange), Ondřejov (510 m a.s.l., green). \textit{Left}: mono (SST-1M-1- dashed line, SST-1M-2- solid line), \textit{Right}: stereo.}
\label{fig:ang}
\end{figure}

Angular resolution for both mono and stereo mode is shown in Figure~\ref{fig:ang} and clearly reflects the expected trend. In the mono regime, improvement is present for energies up to $\sim$ 80\,TeV for Malarg\"ue. In case of Hanle, improvement compared to Ond\v rejov starts to be present at energies $\lesssim$ 13\,TeV surpassing the Malarg\"ue resolution for energies $\lesssim$ 5 TeV. At the highest energies, degradation is evident for this site, which is practically not the case for Malarg\"ue. In stereo mode, the angular resolution exhibits the improvement for energies $\lesssim$ 3\,TeV for both sites. This trend is more distinctive with the increasing altitude. The best achieved resolutions are $\sim$ 0.077$^{\circ}$ at 8\,TeV at Malarg\"ue and $\sim$ 0.085$^{\circ}$ at 5\,TeV at Hanle. At Malarg\"ue, the angular resolution maintains the similar values as at Ond\v rejov for higher energies, with minor degradation. At Hanle, however, it is significantly deteriorated. This effect could potentially be mitigated by further optimization of the spacing between the telescopes. However, our preliminary study suggests that while increasing the spacing leads to improved angular resolution, it has the drawback of decreasing the effective area.

\begin{figure}[h]
\includegraphics[width=.45\linewidth]{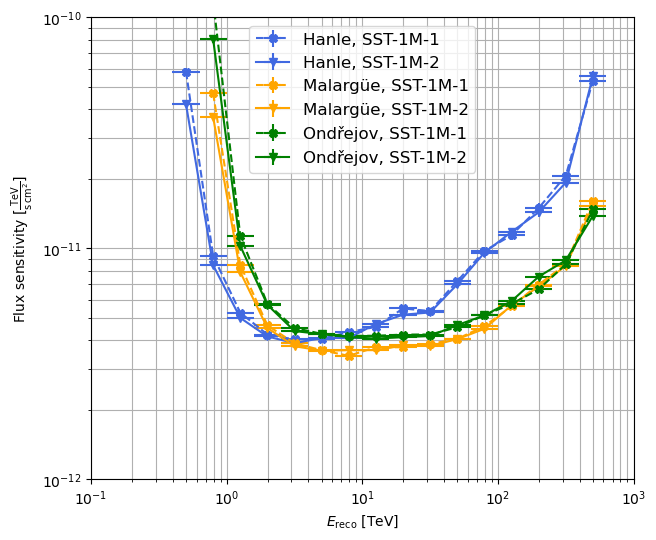}\hfill
\includegraphics[width=.45\linewidth]{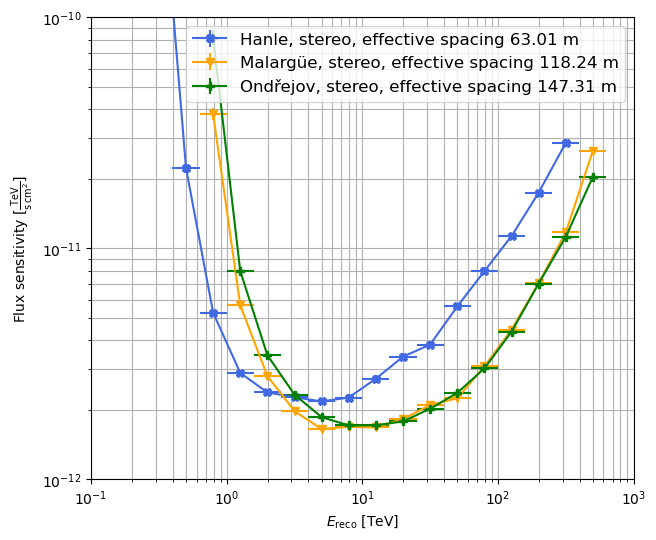}
\caption{Differential sensitivity for different sites: Hanle (4270\,m a.s.l., blue), Malarg\"ue (1420\,m a.s.l., orange), Ond\v rejov (510\,m a.s.l., green). \textit{Left}: mono (SST-1M-1 dashed line, SST-1M-2 solid line), \textit{Right}: stereo.}
\label{fig:sens}
\end{figure}

The preliminary analysis of Receiver Operating Characteristic (ROC), which represents the effectiveness of gamma/hadron ($\gamma$/h) separation, reveals a trend similar to what has been previously discussed: an improvement at lower energies with increasing altitude, at the expense of deterioration at the highest energies.
In Figure ~\ref{fig:sens}, the differential flux sensitivities for both mono and stereo modes are shown. The differential flux sensitivity represents the minimal flux of a source detectable by IACT with at least 5$\sigma$ significance in 50 hours of observation as defined in ~\cite{sensitivity_a, sensitivity_b}. We evaluated the differential sensitivity for on axis point-like gammas re-weighted using the Crab Nebula energy spectrum from~\cite{crab_nebula_spectrum}. Diffuse protons were re-weighted on protons+Helium spectrum~\cite{proton_spectrum}. The behavior of the sensitivities for different sites mirrors the already described trends. In mono mode, the telescopes become more sensitive compared to the Ond\v rejov site for energies $\lesssim$ 80\,TeV at Malarg\"ue site and $\lesssim$ 3\,TeV at Hanle. For energies $\lesssim$ 2\,TeV the telescopes show the best sensitivities at Hanle. However, for this site, the sensitivity is significantly suppressed at higher energies. Similar behavior is observed in the stereo regime, with improved telescope sensitivity for energies $\lesssim$ 5\,TeV at Malarg\"ue and $\lesssim$ 3\,TeV at Hanle compared to Ond\v rejov. Notably, a significant suppression of differential flux sensitivity is observed at Hanle for higher energies with minor deterioration at Malarg\"ue. The time needed to reach 5$\sigma$ detection limit of the Crab Nebula for each site was calculated. The reference values for Ond\v rejov are 1.45$^{+0.08}_{-0.07}$\,h and 0.78$^{+0.05}_{-0.05}$\,h for mono and stereo mode, respectively. For Malarg\"ue, the values are  1.08$^{+0.07}_{-0.05}$\,h for mono mode and 0.62$^{+0.05}_{-0.03}$\,h for stereo mode. For Hanle, the corresponding values are 0.90$^{+0.05}_{-0.05}$\,h for mono mode and 0.47$^{+0.03}_{-0.02}$\,h for stereo mode. The time decreases gradually with increasing site altitude, which is expected due to the improved sensitivity at lower energies. The presented times for mono mode are calculated for SST-1M-1 telescope.

\section{Conclusion}

In this contribution, we presented an analysis of the perfomance of SST-1M telescopes at two sites of different altitudes: Malarg\"ue in Argentina (at an altitude 1420 m a.s.l.) and Hanle in India (at an altitude 4270 m a.s.l.).
Our objective was to study the telescope performance at these different sites compared to their current status. For this purpose, we conducted a detailed MC-simulation based study, which included the atmospheric profiles, geomagnetic field and scaled spacing between the two telescopes for each site. For each site, we analyzed the impact on the effective area, $\gamma$/h separation capabilities, angular and energy resolution and differential flux sensitivity for both mono and stereo modes. An improvement in performance with increasing altitude can be observed at energies lower than few TeV. This is expected as the telescopes are closer to the showers and the attenuation of the Cherenkov light is less significant. This improvement, however, comes at the expense of the degradation of performance at the highest energies. This effect is only minor for the Malarg\"ue site which is $\sim$ 900 m higher than the current site. However it is clearly evident for Hanle, which is significanly higher site with regard to Ond\v rejov.\\
We note that the presented results correspond to the one selected spacing at each site. The complex study of the impact of different spacings on the final performances has not been yet completed and will be the objective of the future work. The impact of telescope layout together with different multiplicities was studied for example in~\cite{cta_layout_config, cta_config_b, cta_australia}. The selected candidate sites are located at different hemispheres, offering access to different astrophysical gamma-ray sources. A study of the possible observation targets was not the objective of this study. We also note that the aspects regarding the logistics for the relocation and maintenance of the telescopes, along with hardware capabilities or adaptation, existing infrastructure, or the number of photometric nights, were not considered in this study, which might also influence the final decision. 

\acknowledgments

\footnotesize{This publication was created as part of the projects funded in Poland by the Minister of Science based on agreements number 2024/WK/03 and DIR/\-WK/2017/12. The construction, calibration, software control and support for operation of the SST-1M cameras is supported by SNF (grants CRSII2\_141877, 20FL21\_154221, CRSII2\_160830, \_166913, 200021-231799), by the Boninchi Foundation and by the Université de Genève, Faculté de Sciences, Département de Physique Nucléaire et Corpusculaire. The Czech partner institutions acknowledge support of the infrastructure and research projects by Ministry of Education, Youth and Sports of the Czech Republic (MEYS) and the European Union funds (EU), MEYS LM2023047, EU/MEYS CZ.02.01.01/00/22\_008/0004632, CZ.02.01.01/00/22\_010/0008598, Co-funded by the European Union (Physics for Future – Grant Agreement No. 101081515), and Czech Science Foundation, GACR 23-05827S.
\\
The authors would like to express their gratitude to Georgios Voutsinas from Université de Genève for providing the atmospheric profile for the Hanle site.}

{\footnotesize
\raggedright
\setlength{\parskip}{0ex}
\setlength{\itemsep}{0pt}
\bibliography{bibfile}{}

\providecommand{\href}[2]{#2}\begingroup\raggedright\begin{thebibliography}{10}

\bibitem{SST-1M}
C.~{Alispach}, A.~{Araudo}, M.~{Balbo}, V.~{Beshley}, A.~{Biland}, J.~{Bla{\v{z}}ek} et~al., \href{https://doi.org/10.1088/1475-7516/2025/02/047}{\emph{JCAP} {\bfseries 2025} (2025) 047} [\href{https://arxiv.org/abs/2409.11310}{{\ttfamily 2409.11310}}].

\bibitem{Alispach_2025b}
C.~Alispach, A.~Araudo, M.~Balbo, V.~Beshley, J.~Bla{\v z}ek, J.~Borkowski et~al., {\emph{Astron. Astrophys.} (2025) }.

\bibitem{veritas_hours}
S.R.P.~for~the VERITAS~collaboration, \href{https://doi.org/10.21468/SciPostPhysProc.13.035}{\emph{SciPost Phys. Proc.} (2023) 035}.

\bibitem{cta_site_selection}
T.~Hassan, L.~Arrabito, K.~Bernlöhr, J.~Bregeon, J.~Cortina, P.~Cumani et~al., \href{https://doi.org/https://doi.org/10.1016/j.astropartphys.2017.05.001}{\emph{Astroparticle Physics} {\bfseries 93} (2017) 76}.

\bibitem{1998cmcc.book.....H}
D.~{Heck}, J.~{Knapp}, J.N.~{Capdevielle}, G.~{Schatz} and T.~{Thouw}, \emph{{CORSIKA: a Monte Carlo code to simulate extensive air showers.}},  Report FZKA 6019, Forschungszentrum Karlsruhe (1998).

\bibitem{urqmd2}
M.~Bleicher et~al., \href{https://doi.org/10.1088/0954-3899/25/9/308}{\emph{J. Phys. G} {\bfseries 25} (1999) 1859} [\href{https://arxiv.org/abs/hep-ph/9909407}{{\ttfamily hep-ph/9909407}}].

\bibitem{qgs}
S.~Ostapchenko, \href{https://doi.org/10.1103/PhysRevD.83.014018}{\emph{Phys. Rev. D} {\bfseries 83} (2011) 014018} [\href{https://arxiv.org/abs/1010.1869}{{\ttfamily 1010.1869}}].

\bibitem{BERNLOHR2008149}
K.~Bernlöhr, \href{https://doi.org/https://doi.org/10.1016/j.astropartphys.2008.07.009}{\emph{Astroparticle Physics} {\bfseries 30} (2008) 149}.

\bibitem{sst1mpipe}
J.~Jurysek, T.~Tavernier, V.~Novotny, P.~Hamal, M.~Heller, J.~Blazek et~al., \emph{sst1mpipe: v0.7.3. 5 february 2025},  Feb., 2025.
\newblock 10.5281/zenodo.14808846.

\bibitem{jurysek_ICRC_proceeding}
J.~{Jurysek}, T.~{Tavernier}, V.~{Novotn{\'y}}, M.~{Heller}, D.~{Mandat}, M.~{Pech} et~al., \emph{{Mono and stereo performance of the two SST-1M telescope prototypes}},  in \emph{38th International Cosmic Ray Conference}, p.~592, Sept., 2024.

\bibitem{crab_nebula_spectrum}
J.~Aleksić, S.~Ansoldi, L.~Antonelli, P.~Antoranz, A.~Babic, P.~Bangale et~al., \href{https://doi.org/https://doi.org/10.1016/j.astropartphys.2015.02.005}{\emph{Astroparticle Physics} {\bfseries 72} (2016) 76}.

\bibitem{sensitivity_a}
H.~Abe, K.~Abe, S.~Abe, A.~Aguasca-Cabot, I.~Agudo, N.~Alvarez~Crespo et~al., \href{https://doi.org/10.3847/1538-4357/ace89d}{\emph{The Astrophysical Journal} {\bfseries 956} (2023) 80}.

\bibitem{sensitivity_b}
{Abe, H.}, {Abe, K.}, {Abe, S.}, {Acciari, V. A.}, {Aguasca-Cabot, A.}, {Agudo, I.} et~al., \href{https://doi.org/10.1051/0004-6361/202346927}{\emph{A\&A} {\bfseries 680} (2023) A66}.

\bibitem{proton_spectrum}
{\scshape DAMPE Collaboration} Collaboration, \href{https://doi.org/10.1103/PhysRevD.109.L121101}{\emph{Phys. Rev. D} {\bfseries 109} (2024) L121101}.

\bibitem{cta_layout_config}
A.~Acharyya, I.~Agudo, E.~Angüner, R.~Alfaro, J.~Alfaro, C.~Alispach et~al., \href{https://doi.org/https://doi.org/10.1016/j.astropartphys.2019.04.001}{\emph{Astroparticle Physics} {\bfseries 111} (2019) 35}.

\bibitem{cta_config_b}
K.~Bernlöhr, A.~Barnacka, Y.~Becherini, O.~{Blanch Bigas}, E.~Carmona, P.~Colin et~al., \href{https://doi.org/https://doi.org/10.1016/j.astropartphys.2012.10.002}{\emph{Astroparticle Physics} {\bfseries 43} (2013) 171}.

\bibitem{cta_australia}
S.~Lee, S.~Einecke, G.~Rowell, C.~Balazs, J.A.~Bellido, S.~Dai et~al., \href{https://doi.org/10.1017/pasa.2022.29}{\emph{Publications of the Astronomical Society of Australia} {\bfseries 39} (2022) e041}.

\end{thebibliography}\endgroup


\begin{thebibliography}{99}
\bibitem{Alispach_2025a} C. Alispach et al. \href{https://doi.org/10.1088/1475-7516/2025/02/047}{JCAP02(2025)047}
\bibitem{icrc23_performance} Juryšek et al. \href{https://doi.org/10.22323/1.444.0592}{PoS(ICRC2023)592}
\bibitem{Alispach_2025b} C. Alispach et al. \href{https://doi.org/10.1051/0004-6361/202555292}{Astronomy \& Astrophysics (2025)}
\end{thebibliography}
\bibliographystyle{JHEP_mod}
}

\clearpage
\section*{Full Authors List: SST-1M Collaboration}
\scriptsize
\noindent
C.~Alispach$^1$,
A.~Araudo$^2$,
M.~Balbo$^1$,
V.~Beshley$^3$,
J.~Bla\v{z}ek$^2$,
J.~Borkowski$^4$,
S.~Boula$^5$,
T.~Bulik$^6$,
F.~Cadoux$^`$,
S.~Casanova$^5$,
A.~Christov$^2$,
J.~Chudoba$^2$,
L.~Chytka$^7$,
P.~\v{C}echvala$^2$,
P.~D\v{e}dic$^2$,
D.~della Volpe$^1$,
Y.~Favre$^1$,
M.~Garczarczyk$^8$,
L.~Gibaud$^9$,
T.~Gieras$^5$,
E.~G{\l}owacki$^9$,
P.~Hamal$^7$,
M.~Heller$^1$,
M.~Hrabovsk\'y$^7$,
P.~Jane\v{c}ek$^2$,
M.~Jel\'inek$^{10}$,
V.~J\'ilek$^7$,
J.~Jury\v{s}ek$^2$,
V.~Karas$^{11}$,
B.~Lacave$^1$,
E.~Lyard$^{12}$,
E.~Mach$^5$,
D.~Mand\'at$^2$,
W.~Marek$^5$,
S.~Michal$^7$,
J.~Micha{\l}owski$^5$,
M.~Miro\'n$^9$,
R.~Moderski$^4$,
T.~Montaruli$^1$,
A.~Muraczewski$^4$,
S.~R.~Muthyala$^2$,
A.~L.~Müller$^2$,
A.~Nagai$^1$,
K.~Nalewajski$^5$,
D.~Neise$^{13}$,
J.~Niemiec$^5$,
M.~Niko{\l}ajuk$^9$,
V.~Novotn\'y$^{2,14}$,
M.~Ostrowski$^{15}$,
M.~Palatka$^2$,
M.~Pech$^2$,
M.~Prouza$^2$,
P.~Schovanek$^2$,
V.~Sliusar$^{12}$,
{\L}.~Stawarz$^{15}$,
R.~Sternberger$^8$,
M.~Stodulska$^1$,
J.~\'{S}wierblewski$^5$,
P.~\'{S}wierk$^5$,
J.~\v{S}trobl$^{10}$,
T.~Tavernier$^2$,
P.~Tr\'avn\'i\v{c}ek$^2$,
I.~Troyano Pujadas$^1$,
J.~V\'icha$^2$,
R.~Walter$^{12}$,
K.~Zi{\c e}tara$^{15}$ \\

\noindent
$^1$D\'epartement de Physique Nucl\'eaire, Facult\'e de Sciences, Universit\'e de Gen\`eve, 24 Quai Ernest Ansermet, CH-1205 Gen\`eve, Switzerland.
$^2$FZU - Institute of Physics of the Czech Academy of Sciences, Na Slovance 1999/2, Prague 8, Czech Republic.
$^3$Pidstryhach Institute for Applied Problems of Mechanics and Mathematics, National Academy of Sciences of Ukraine, 3-b Naukova St., 79060, Lviv, Ukraine.
$^4$Nicolaus Copernicus Astronomical Center, Polish Academy of Sciences, ul. Bartycka 18, 00-716 Warsaw, Poland.
$^5$Institute of Nuclear Physics, Polish Academy of Sciences, PL-31342 Krakow, Poland.
$^6$Astronomical Observatory, University of Warsaw, Al. Ujazdowskie 4, 00-478 Warsaw, Poland.
$^7$Palack\'y University Olomouc, Faculty of Science, 17. listopadu 50, Olomouc, Czech Republic.
$^8$Deutsches Elektronen-Synchrotron (DESY) Platanenallee 6, D-15738 Zeuthen, Germany.
$^9$Faculty of Physics, University of Bia{\l}ystok, ul. K. Cio{\l}kowskiego 1L, 15-245 Bia{\l}ystok, Poland.
$^{10}$Astronomical Institute of the Czech Academy of Sciences, Fri\v{c}ova~298, CZ-25165 Ond\v{r}ejov, Czech Republic.
$^{11}$Astronomical Institute of the Czech Academy of Sciences, Bo\v{c}n\'i~II 1401, CZ-14100 Prague, Czech Republic.
$^{12}$D\'epartement d'Astronomie, Facult\'e de Science, Universit\'e de Gen\`eve, Chemin d'Ecogia 16, CH-1290 Versoix, Switzerland.
$^{13}$ETH Zurich, Institute for Particle Physics and Astrophysics, Otto-Stern-Weg 5, 8093 Zurich, Switzerland.
$^{14}$Institute of Particle and Nuclear Physics, Faculty of Mathematics and Physics, Charles University, V Hole\v sovi\v ck\' ach 2, Prague 8, Czech~Republic.
$^{15}$Astronomical Observatory, Jagiellonian University, ul. Orla 171, 30-244 Krakow, Poland.

\end{document}